\documentclass[aps,pre,reprint,twoside,showkeys,longbibliography,superscriptaddress]{revtex4-2}

\usepackage{graphicx}        
\usepackage{amsmath,amssymb,amsfonts}
\usepackage{bm}              
\usepackage{bbm}             
\usepackage{booktabs}        
\usepackage[inline,shortlabels]{enumitem}
\usepackage[dvipsnames]{xcolor}
\usepackage[colorlinks=true,linkcolor=blue,citecolor=blue,urlcolor=Salmon]{hyperref}
\usepackage{url}
\usepackage[normalem]{ulem}  
\usepackage{placeins} 
\usepackage{float}

\newcommand{\mat}[1]{\mathbf{#1}}

\begin{document}

\title{Scale invariance and statistical significance in complex weighted networks}

\author{Filipi N. Silva}
\affiliation{Center for Complex Networks and Systems Research, Luddy School of Informatics, Computing, and Engineering, Indiana University Bloomington, USA}

\author{Sadamori Kojaku}
\affiliation{Department of Systems Science and Industrial Engineering, Binghamton University, Binghamton, New York, USA}

\author{Alessandro Flammini}
\affiliation{Center for Complex Networks and Systems Research, Luddy School of Informatics, Computing, and Engineering, Indiana University Bloomington, USA}

\author{Filippo Radicchi}
\affiliation{Center for Complex Networks and Systems Research, Luddy School of Informatics, Computing, and Engineering, Indiana University Bloomington, USA}

\author{Santo Fortunato}
\affiliation{Center for Complex Networks and Systems Research, Luddy School of Informatics, Computing, and Engineering, Indiana University Bloomington, USA}

\date{\today} 

\begin{abstract}
Most networks encountered in nature, society, and technology have weighted edges, representing the strength of the interaction/association between their vertices. Randomizing the structure of a network is a classic procedure used to estimate the statistical significance of properties of the network, such as transitivity, centrality and community structure.
Randomization of weighted networks has traditionally been done via the weighted configuration model (WCM),
a simple extension of the configuration model, where weights are interpreted as bundles of edges.
It has previously been shown that the ensemble of randomizations provided by the WCM is affected by the specific scale used to compute the weights, but the consequences for statistical significance were unclear. Here we find that statistical significance based on the WCM is scale-dependent, whereas in most cases results should be independent of the choice of the scale. More generally, we find that designing a null model that does not violate scale invariance is challenging. A two-step approach, originally introduced for network reconstruction, in which one first randomizes the structure, then the weights, with a suitable distribution, restores scale invariance, and allows us to conduct unbiased assessments of significance on weighted networks.
\end{abstract}

\maketitle

\section{Introduction}

Networks are regularly used to represent any sort of system. From the first sociograms, used to visualize the social interactions between pupils in a classroom~\cite{moreno34}, to the Internet, the brain, biological networks, information networks, transportation networks, and financial networks, scholars have learned a lot by reducing systems to the set of their elementary units (vertices) and their mutual interactions (edges)~\cite{newman10,barabasi16,menczer20}.

Networks representing real systems are a mix of order and randomness. The order comes from the processes that generate the edges between the vertices, which can often be reduced to a simple set of rules. The randomness comes from the stochastic character of such processes, due to uncontrollable factors that play a role in the final placement of the edges. If real networks were entirely random, they would not be very interesting. This is why it is important to
single out their random component, via statistical hypothesis testing.  The latter requires null models, i.e., sets of rules that generate randomized versions of the network under investigation, that maximize the disorder in its structure, under some constraints. The most popular null model is the Configuration Model (CM)~\cite{molloy95,fosdick2018configuring,chung02}, where the edges of the network are randomly repositioned, such to preserve the degree of each vertex, which is the number of vertices it is connected to (neighbors). The choice of this constraint is motivated by the great importance that degree has in the structure and function of
real networks~\cite{newman10,barabasi16,menczer20}. In this way, if a property of the network does not appear in the corresponding CM randomizations, one can claim that that property is not simply due to the degrees of the vertices. Quantitatively, this is done by computing the $p$-value of the score calculated on the original network, with respect to the distribution of scores in its null model randomizations. If the $p$-value is sufficiently small (typically below 0.05), then we can argue that the score is statistically
significant. This applies to any variable that can be computed on the network. For instance, the average clustering coefficient estimates the average local density of triangles of the network~\cite{watts98}. If the average clustering coefficient of a given network is statistically significant with respect to its CM randomizations, the observed local density of triangles cannot be reproduced simply because vertices have certain degrees, but there must be another mechanism at play.  This framework is used in motif analysis~\cite{milo2002network,sporns2004motifs}, community detection~\cite{ruan08,spirin03,kojaku2018generalised,in_t_veld2014finding}, and network sparsification~\cite{gemmetto2017irreducible}.

Real networks are often weighted, in that their edges carry a value expressing how strongly the corresponding pairs of vertices are interacting or associated~\cite{barrat04}.
For instance, in a social network, weights may represent the number or duration of social interactions between individuals. To assess the statistical significance of measures on weighted networks, we then need to define how to randomize their structure. The simplest recipe is to assume that a weighted edge is a multi-edge, i.e., a bundle of elementary edges
with weight one~\cite{newman04}, and to reposition the elementary edges exactly like the CM does.
In this case, what is preserved is the \textit{weighted degree}, or \textit{strength}, of each vertex, i.e., the sum of the weights of all edges attached to it. This procedure is natural if the weights are integers. If they have real values it was suggested that one could multiply them by a sufficiently large constant $A$, such that all weights become integers (with good approximation), turn them into multi-edges, do the CM randomization, and divide the final weights by $A$. This is the Weighted Configuration Model (WCM)~\cite{newman04,serrano05}.

In many instances the scale adopted to express the weights is immaterial. In a social network, where weights are the duration of personal interactions between people, it should not matter whether time is measured in seconds, minutes or hours. Likewise, if we consider the World Trade Web (WTW)~\cite{serrano03}, where vertices are countries and edges represent their trade relationships, it does not matter whether the trade flows are expressed in thousands, millions, or billions of US dollars. Indeed, when one inspects the mathematical expression of key scores typically computed on networks, the choice of the scale does not matter, as we shall see. Previous work has shown that there is a non-trivial scale dependence of the network randomizations generated by the WCM~\cite{garlaschelli09,mastrandrea14}, but the consequences of this finding have not been explored. 
In this paper we show that, when estimating statistical significance using the WCM as null model, the final assessment depends on the choice of the scale, against intuition. The reason lies in the peculiar features of the  distribution of the weight of any given edge generated by the CM and, consequently, by the WCM. A two-step null model, previously introduced for network reconstruction~\cite{parisi20}, in which one first randomizes the placement of the edges and then assigns weights to the edges by extracting them from a certain distribution, restores scale invariance.

\section{Results}\label{sec:result}

\subsection{Network measures and scale invariance}
\label{sec:measures}

Here we show that traditional measures defined on weighted networks are not dependent of the choice of the weight scale. Let us suppose to have a weighted network $G$ with $n$ vertices and $m$ edges. ${\bf W}$ is the $n\times n$ \textit{weight matrix}, whose entry $W_{ij}$ indicates the weight of the edge joining vertices $i$ and $j$ (if there is no edge $W_{ij}=0$). We define a \textit{rescaling} by multiplying all elements of ${\bf W}$ by a constant $A>0$. This leads to a matrix ${\bf W^\prime}=A{\bf W}$.
Next, we will check how different network variables change after this transformation.

The strength of vertex $i$ is defined as

\begin{equation}
s_{i}=\sum_{j}W_{ij}\,.
    \label{eq:strength}
\end{equation}

When the edge weights are scaled by a factor $A$, the strength of vertex $i$ is also scaled by the same factor, becoming

\begin{equation}
s^\prime_{i}=\sum_{j}W^\prime_{ij}=A\sum_{j}W_{ij}=As_i. \label{eq:strength1}
\end{equation}

Note that the change of scale is immaterial, i.e., the scaling does not alters neither the ranking of the vertices based on strength, nor the strength distribution. Scaling does not affect many network measures. For example, the weighted clustering coefficient of vertex $i$, in the formulation by Onnela et al.~\cite{onnela05}, reads

\begin{equation}
C_{i}=\frac{2}{k_i(k_i-1)}\sum_{jk}\left(\tilde{W}_{ij}\tilde{W}_{jk}\tilde{W}_{ki}\right),
    \label{eq:wclus_coeff}
\end{equation}
where $k_i$ is the degree of vertex $i$, $\tilde{W}_{ij}=W_{ij}/W_{max}$, $W_{max}$ being the largest edge weight of the network.
Since $\tilde{W}_{ij}$ is dimensionless, it is not affected by any change of scale, so ${C}^\prime_i=C_i$.

The eigenvector centrality~\cite{bonacich87} is also scale invariant.
The eigenvector centrality of vertex $i$ is the $i$-th entry of the principal eigenvector of the adjacency matrix.
The extension to weighted networks simply involves using the weight matrix instead of the adjacency matrix, and the scaling of edge weights scale the eigenvalues while leaving the eigenvectors intact.

%
%
%
%
%

Finally, we consider community structure, i.e., the peculiar organization of many real networks into groups of vertices, called communities, clusters, or modules, with a comparatively higher density of edges within the groups than between them~\cite{porter09, fortunato10,fortunato16,fortunato22}. The most popular method to detect communities in networks is maximizing modularity, a quality function that expresses the goodness of a division into clusters~\cite{newman04b}. Modularity can be easily extended to the weighted case.
The weighted modularity of a partition ${\bf g}$ of the network into communities is~\cite{newman04}

\begin{equation}
Q=\frac{1}{2W}\sum_{ij} \left(W_{ij}-\frac{s_is_j}{2W}\right)\delta_{g_ig_j},
\label{eq:modul}
\end{equation}
where $W$ is the total weight on the edges, $s_i$ ($s_j$) is the strength of vertex $i$ ($j$), $g_i$ ($g_j$) is the community label of $i$ ($j$), and $\delta_{g_ig_j}$ the Kronecker delta, which yields one when $i$ and $j$ are in the same community (same label, i.e., $g_i = g_j$) and zero otherwise.
Since both the numerator and denominator are linear with respect to the scale factor $A$, the weighted modularity is scale invariant.
Consequently, a change of scale does not change the measure: any partition will have the same value of $Q$ for any choice of the scale. In particular, the partition with largest modularity, which is supposed to be the best one, will be the same and have the same value of $Q$ regardless of $A$.

\subsection{Statistical significance and scale invariance}
\label{sec:stat_sign_scal}
\begin{figure}
    \centering
    \includegraphics[width=0.95\columnwidth]{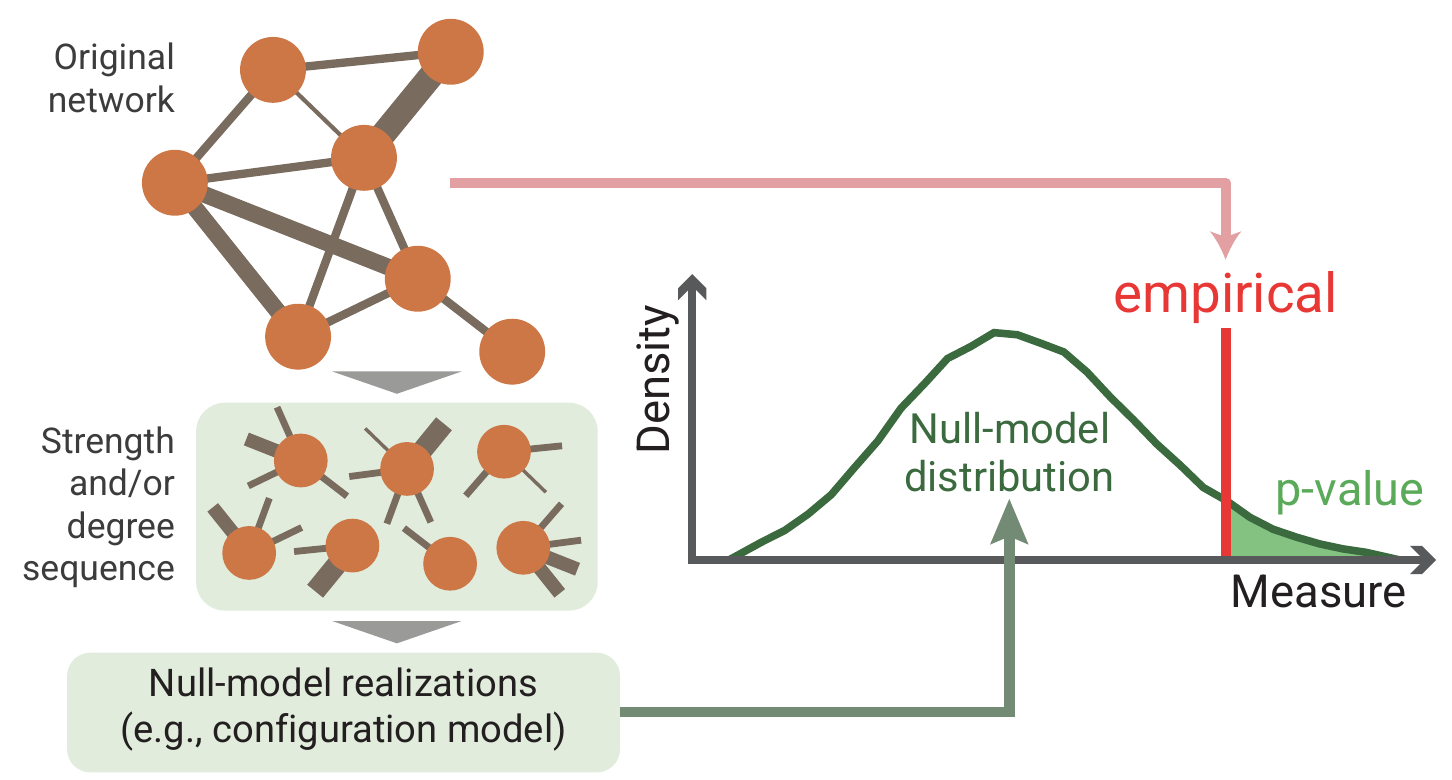}
    \caption{Statistical significance of network variables. The $p$-value of the chosen variable measured on the original network equals the area under the curve of the null model distribution of the variable, to the right of the measured value if we argue that it is higher than on randomized networks (to the left if it is lower).}
    \label{fig:p-value}
\end{figure}

Let us consider a network $G$ with weight matrix ${\bf W}$ and a variable $F$ defined on $G$. We indicate with $F({\bf W})$ the value of $F$ on $G$, via its weight matrix ${\bf W}$. To determine the statistical significance of $F({\bf W})$, one calculates $F$ on a random sample of the randomizations of $G$, according to the chosen null model, and compute the $p$-value of $F({\bf W})$ with respect to the distribution of $F$ on the randomizations. The procedure is schematically illustrated in Fig.~\ref{fig:p-value}.

For the CM random networks are constructed as follows.
We first break each edge into two halves, called stubs, which generates $2m$ stubs for a network with $m$ edges.
Then we combine pairs of stubs at random, until there are no stubs available.
The probability of picking a stub attached at $i$ among the $2m$ stubs is $k_i/2m$, where $k_i$ is the degree of vertex $i$. The probability of picking another stub attached at $j$ among the remaining $2m-1$ stubs is $k_j/(2m-1)$.
Therefore, the probability of pairing $i$ and $j$ is $(k_i/2m) \times k_j / (2m-1)\sim k_ik_j/4m^2$.
The distribution of the number of edges connecting $i$ and $j$ over all possible randomizations follows a hypergeometric distribution with mean $k_ik_j/2m$ (see Appendix~\ref{appendix:weighted_chunglu_model}).

In the case of weighted networks, the degrees are replaced by the strength $s_i$ and $s_j$ and the total number $m$ of edges by the total edge weight $W$.
If we interpret an edge weight $W_{ij}$ as a multi-edge consisting of $W_{ij}$ elementary edges with weight one, the procedure that we have described above for the CM can be naturally extended to weighted networks. As a result, the expected weight of the edge joining $i$ and $j$ in the randomizations generated by the WCM is $s_is_j/2W$, and the distribution of the weights is hypergeometric, assuming that the weights of the original network are integers. If the weights are not integers, one can discretize them by multiplying them by a large enough factor $A$, so that they become integers with good approximation. Then one can operate as in the case of integer weights, with the additional final step of dividing the weights produced by the randomization by $A$~\cite{newman04}. After the discretization, the distribution of weights is still hypergeometric with mean
$A s_is_j/2W$.

For statistical significance to be invariant with respect to the choice of the weight scale, the null model distribution of the chosen variable $F$ must be independent of the scale factor $A$. Let us indicate with ${\bf W^R}$ the weight matrix of a generic randomization generated by the WCM. We can write

\begin{equation}
W_{ij}^R=\frac{s_is_j}{2W}+\delta W^R_{ij},
\label{eq:mean}
\end{equation}

where $\delta W^R_{ij}$ is the fluctuation with respect to the expected value $W^{\text{avg}}_{ij}=s_is_j/2W$ of the weight of the edge $ij$.
Most network variables are implicit or explicit functions $F({\bf W})=F(W_{12}, W_{13}, \dots, W_{1n})$ of the edge weights. If we do a first-order expansion of $F({\bf W^R})$ around the expected values ${\bf W^{avg}}$ we have

\begin{equation}
F({\bf W^R})\approx F({\bf W^{avg}})+\sum_{i< j}\left.{\frac{\partial F}{\partial W_{ij}}}\right|_{W_{ij}=s_is_j/2W}\delta W^R_{ij}.
\label{eq:expansion}
\end{equation}

If the function $F$ is dimensionless in the weights and a function of edge weights, like the weighted clustering coefficient, maximum eigenvector centrality, and weighted modularity, which
we have introduced in Section~\ref{sec:measures}, their derivatives have the dimension of an inverse weight, so they scale as $A^{-1}$. The fluctuations $\delta W^R_{ij}$, instead, are proportional to the standard deviation of a hypergeometric distribution with mean $s_is_j/2W$, which equals the square root of the mean $\sqrt{s_is_j/2W}$ with good approximation (provided $s_i\ll W$, $\forall i$). Consequently, the variation of
$F({\bf W^R})$ around the expected value $F({\bf W^{avg}})$
scales as $A^{-1/2}$. Hence, the null model distribution of $F$ shrinks as the scale factor increases and the statistical significant assessment depends on the weight scale. In the limit of large $A$, the width of the distribution goes to zero, which implies that any value of the variable would be statistically significant.

\begin{figure*}
    \centering
    \includegraphics[width=0.98\textwidth]{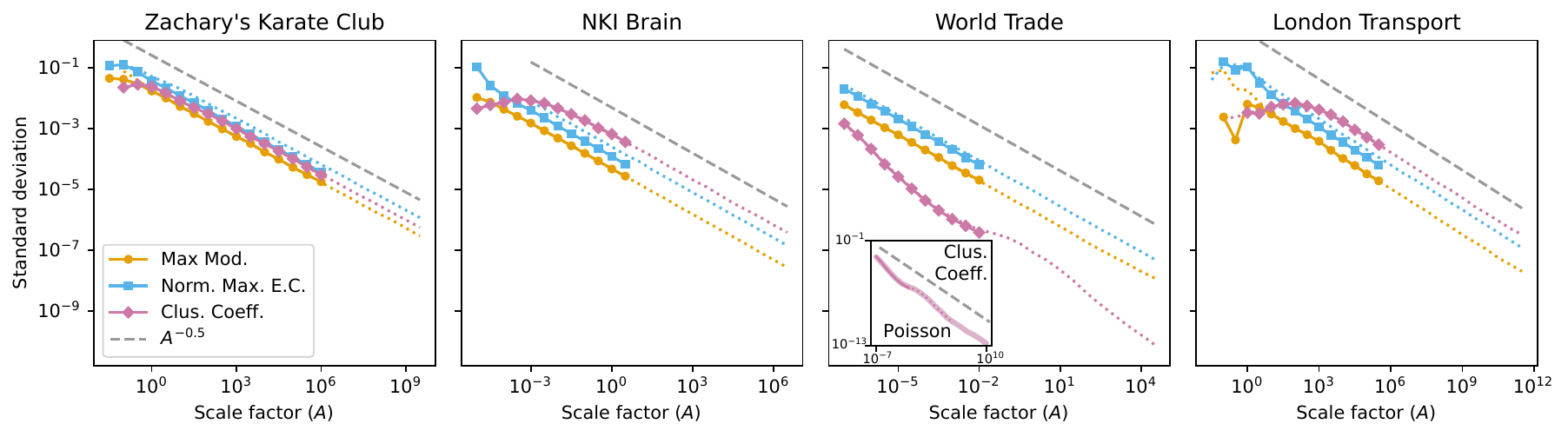}
    \caption{Scale dependence of the standard deviation of the WCM distribution for the weighted clustering coefficient, maximum eigenvector centrality, and maximum modularity in four networks: Zachary's karate club, NKI Brain, World Trade, and London Transport (see Section~\ref{subsec:data} for details). The dashed line represents the conjectured inverse square root behavior, which the three curves follow closely in each case as $A$ increases. Dotted lines indicate results obtained from the weighted Chung–Lu model using hypergeometric distributions (see section~\ref{subsec:wchunglu}). The inset for the World Trade network shows the Poisson approximation for large $A$-values of the WCM model for the weighted clustering coefficient.}
    \label{fig:Ascaling}
\end{figure*}

In Fig.~\ref{fig:Ascaling} we show how the standard deviation of the WCM distribution varies as a function of $A$ for the weighted clustering coefficient, the maximum eigenvector centrality, and the maximum modularity, for four real networks (described in the Methods). Due to the large number of stubs generated for large $A$, for the asymptotic behavior of the curves (dotted lines) we adopted a Weighted Chung-Lu Model (see Section~\ref{subsec:wchunglu} for details), which is the canonical version of the WCM, where the strengths of the vertices are preserved only in expectation. The standard deviation indeed decreases as the inverse square root of $A$, for large enough $A$.

We stress that the derivatives in Eq.~\ref{eq:expansion} do not depend on
${\bf W^R}$. For weighted modularity this is true if we compute it on any fixed partition ${\bf g}$. But if we maximize the measure over all partitions, to find the best division in communities, we generally obtain different partitions, so the functional form of $F$ changes with ${\bf W^R}$ and its derivatives are then dependent of the weight assignments of the specific randomization. However, our argument in Section~\ref{sec:stat_sign_scal} is still valid.
In addition, for weighted modularity, $F({\bf W^{avg}})=0$, because the expected value of $W^R_{ij}$ exactly matches the term $s_is_j/2W$ in Eq.~\ref{eq:modul}. Hence, the expected value of the maximum modularity itself tends to zero as $A$ goes to infinity, following the same $A^{-1/2}$ decay of its standard deviation.

\subsection{Which null model?}
\label{sec:newnull}

Our analysis shows that a widely popular null model, the WCM, which has been used for two decades, cannot provide a reliable assessment of the statistical significance of network metrics. The reason lies in the supposed equivalence between weights and multi-edges. The moment we establish a discrete scale, the WCM generates weighted networks where the size of the fluctuations around the expected edge weight values depends on the chosen scale.

The dependence on scale arises because the random fluctuations in the WCM do not increase in proportion to the expected value as the scale factor $A$ changes.
One might think to resolve this by ensuring that the fluctuations $\delta W^R_{ij}$ scale linearly with $A$, for example by drawing the weights $W_{ij}$ from an exponential distribution with mean $s_is_j/2W$, since both the mean and standard deviation of the exponential scale together.
Although this approach addresses the issue, the resulting random networks are fully connected, which does not reflect the sparsity of real-world networks.
This is because the exponential distribution almost never yields exactly zero value of edge weights, leading to random networks that are fully connected, with every possible edge present.

Another issue of the exponential distribution, along with the WCM, is that they do not preserve the degree sequence of the network. In particular, the degree sequence of a weighted network can be quite different from the expected one on randomizations generated by the WCM~\cite{fagiolo13,squartini11}. This suggests that degrees and strengths are irreducible variables and that
one cannot use either set to surrogate the information given by the other~\cite{mastrandrea14}.
Because the sequences of the vertex degrees and strengths have different functional roles in determining or constraining the structure of a network, scholars have developed null models, where both the degree and the strength sequence are constrained~\cite{garlaschelli09,serrano06, 
bianconi09,mastrandrea14}, generating the most informative network randomizations based on degree and strength sequences. In the Enhanced Configuration Model (ECM),
the expected values $\langle k_i\rangle$ and $\langle s_i\rangle$ of the degree and the strength of every vertex, coincide with the actual values $k_i$ and $s_i$ of the original network~\cite{mastrandrea14}. The ECM works for integer weights, but has also been extended to the continuous case~\cite{gabrielli19,parisi20}. However, the model constrains degrees and strengths together and, consequently, the parameters controlling the edge weights also play a role in determining the connection probabilities. The Separable Enhanced Configuration Model (SECM)~\cite{gabrielli19}, instead, constrains degrees and strength separately, following a two-step procedure, where first one generates the structure of the network based on the degree sequence, which is preserved in expectation (canonical ensemble), and then one generates the weights on the edges of the resulting configurations, such that the strength sequence is preserved, again in expectation. For any given edge, the weight is extracted from an exponential (maximum entropy) distribution. Generalized versions of the SECM, CReMa and CReMb, have been successively proposed, by using as input arbitrary probability distributions for the adjacency matrices~\cite{parisi20}. The SECM, CReMa, and CReMb are defined such that a change in the scale of weights is entirely reabsorbed in the Lagrange multipliers used to account for the constraints, so that the expectation values of variables are invariant upon a change of units. 

\begin{figure*}
    \centering
    \includegraphics[width=0.98\textwidth]{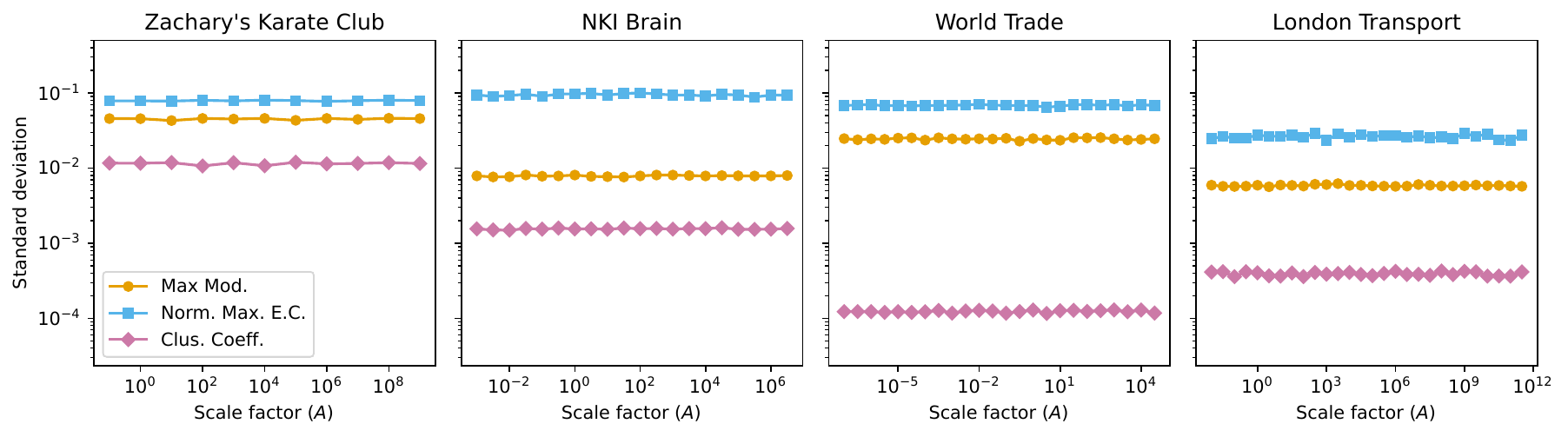}
    \caption{Scale dependence of the distribution of our variant of the CReMb, that separates structure and weights. The variables and the networks are the same as in Fig.~\ref{fig:Ascaling}. The curves are approximately flat, signaling scale invariance.}
    \label{fig:SECM_figs}
\end{figure*}

While there are nuances distinguishing the various models and a certain freedom in choosing one or another, here we consider a variant of the CReMb and verify that statistical significance is scale invariant according to this variant.
The key difference is that the CReMb specifies the probability of an edge based on the density-corrected Gravity Model~\cite{cimini2015systemic}, while our variant uses the configuration model, so it is much faster.
For simplicity, we assume that the expected value of the weight of the edge between vertices $i$ and $j$ is $ms_is_j/(Wk_ik_j)$.
Hence, after generating the structure of the network with the CM (or its canonical version~\cite{chung02}, where degrees are preserved only in expectation), the weight of edge $ij$ (if present) is extracted from an exponential distribution with mean $ms_is_j/(Wk_ik_j)$ (see Section~\ref{subsec:SECM} for details). 
In Fig.~\ref{fig:SECM_figs} we show how the standard deviation of the null model distribution varies with the scale factor $A$, for the same four networks featured in Fig.~\ref{fig:Ascaling}. The flat pattern shows that there is no dependence on $A$ and that, therefore,
statistical significance does not depend on the choice of the scale, as it should be.

\begin{figure*}
    \centering
    \includegraphics[width=0.95\textwidth]{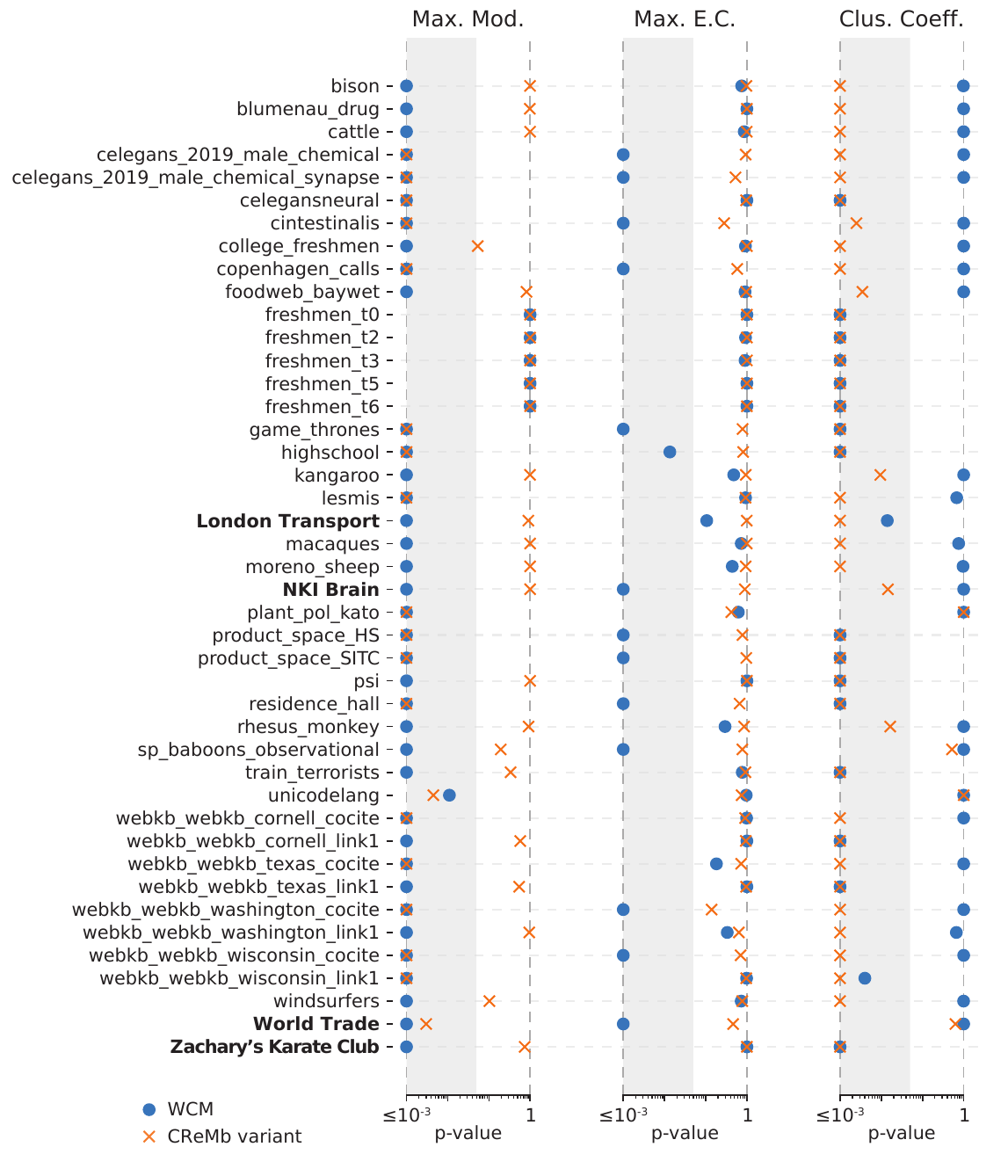}
    \caption{Statistical significance of the three focal variables for the WCM and our variant of the CReMb on a large collection of real networks. For each network we indicate the $p$-value of each variable (dots for the WCM, crosses for the CReMb). Since the WCM is not scale invariant, we compute the $p$-values considering the original weights of the networks, without rescaling. The shaded regions correspond to $p$-values $\leq 0.05$.}
    \label{fig:table}
\end{figure*}

In Fig.~\ref{fig:table} we assess the significance of the values of the three variables on many networks, according to the WCM and our CReMb variant.
For our model we see that the weighted clustering coefficient is significant in most cases, whereas the maximum eigenvector centrality is almost always not significant. This is due to the fact that eigenvector centrality is strongly correlated to the strength sequence of the network, which is kept (approximately) fixed by the randomization. The modularity maximum, instead, is significant or not depending on the network.


\section{Discussion}
\label{sec:disc}

Assessing the statistical significance of network measures is the natural procedure to estimate the degree of randomness in the structure of the network. It requires a null model generating randomizations of the network. Randomizing the structure simply means repositioning the edges at random, while preserving some constraints, and it is widely established in network science.

In order to randomize a weighted network one needs to specify how to shuffle the edges but also how to assign weights to them.
The traditional prescription is the one of the WCM, where weighted edges are considered as bundles of elementary edges, which can then be randomly repositioned as it is done for unweighted networks, and finally grouped together to generate the weights. In this paper, we have shown that the WCM is not reliable to assess statistical significance of network variables, because $p$-values depend on the specific scale adopted to measure the weights, against the expectation that the significance of network measures should not depend on the scale. 

The problem stems from the fact that the number of elementary edges (weights) between any two vertices in the WCM follows a hypergeometric distribution, whose mean and standard deviation have different dependencies on the scale. For the exponential distribution, instead, they both grow linearly with the scale; hence, the exponential offers a natural scale-invariant alternative.
However, using the exponential or any other continuous distribution to assign the weight to any edge has a major drawback, in that every pair of vertices would end up being connected, which is unusual for most real-world networks.

A simple way out consists of separating the structure of the network from the edge weights. This way, one could first randomize the structure, by reshuffling the edges of the original network, and then assign weights to them, extracted from some scale-invariant distribution, like the exponential above. This approach had already been proposed in the context of network reconstruction, where the exponential was naturally identified as the maximum-entropy distribution given the weighted constraints and the (separate) information on the purely topological null model~\cite{parisi20}. We showed that by doing that the $p$-values obtained for three selected weighted network measures, i.e., the clustering coefficient, the maximum eigenvector centrality, and the maximum modularity, do not depend on the weight scale, providing a reliable assessment of their significance.
In particular situations, in which a specific weight scale is naturally identifiable, the WCM could still be used.

A key lesson from this work is that
finding a null model for weighted networks is non-trivial and that scale invariance, which should be postulated from any credible null model, strongly constrains the set of possible choices. Another lesson is that structure and weight, which do not appear to mix well, should be naturally treated as separate entities. Finally, since the expression of weighted modularity (Eq.~\ref{eq:modul}) was derived having the WCM in mind, one might be tempted to conclude that it has to be changed. However, it only uses the mean value of the null model distribution, leaving aside any other feature. Hence, any null model based on the same mean value would lead to the same expression of modularity. The latter then does not need to be modified, only reinterpreted, although the underlying model has to be scale-invariant.

\section{Methods}
\subsection{Data}
\label{subsec:data}

To illustrate the relationship between the considered measures in the null model and the scale factor (Figs.~\ref{fig:Ascaling}, \ref{fig:SECM_figs}), we employed four real-world weighted networks, from different domains:

{\bf Zachary's Karate Club}: A weighted version of the well-known social network originally introduced by Zachary~\cite{zachary77}. Vertices represent members of a university karate club, and edges reflect social interactions between them. In the weighted variant used here, the edge weights represent the number of social contexts in which pairs of members interacted.

{\bf NKI Brain}: Structural brain network from the enhanced NKI–Rockland lifespan sample (eNKI‑RS)~\cite{nooner2012nki,faskowitz2018wsbm}, which provides $200 \times 200$ structural connectivity (SC) matrices for 558 subjects. The weights encode the strength of white‑matter connections between cortical and subcortical regions. Only the SC version of the dataset was used. The network analyzed in this study corresponds to subject number 50.

{\bf World Trade}: A network of trading interactions between the 20 countries with the highest trade volume in 2019, spanning from 1988 onward. The network was constructed using data from the World Bank's Trade and Tariff Data~\cite{vargas2024open}. Each vertex represents a country, and weighted edges correspond to the annual import trade value (in thousands of US dollars) between country pairs.

{\bf London Transport}: A transportation network of London~\cite{DeDomenico2014Navigability}. Vertices correspond to train, tube, and DLR stations. An edge connects two stations if there is a service stopping at both. Weights represent the estimated travel time between stations, derived from average speeds and include estimated transfer costs between layers of transport.

To understand the impact of the choice of null models on the estimated $p$-values for the various network measures, we additionally considered a diverse collection of networks obtained from the Netzschleuder dataset~\cite{Peixoto2020Netzschleuder}. A brief description of the networks can be found in Table~\ref{tab:networks}.

\begin{table*}
\centering
\caption{Description of the networks used in Figure~\ref{fig:table}.
All networks are available in the Netzschleuder dataset~\cite{Peixoto2020Netzschleuder},
with their names listed in the \textit{Identifier} column.
The table also specifies the meaning of the edge weights for each network.}
\label{tab:networks}
\footnotesize
\begin{ruledtabular}
\begin{tabular}{r|ll}

\textbf{Identifier}                        & \textbf{Description}                     & \textbf{Weight} \\
  \colrule
bison                                   & Dominance network of bison               & Dominance events              \\
blumenau\_drug                          & Drug interaction network Blumenau        & Interaction severity          \\
cattle                                  & Dominance network of cattle              & Dominance frequency           \\
celegans\_2019\_male\_chemical          & C. elegans male chemical connectome      & Connectivity from EM sections \\
celegans\_2019\_male\_chemical\_synapse & Directed C. elegans male connectome      & Number of synapses            \\
celegansneural                          & Classic C. elegans connectome            & Synapse count                 \\
cintestinalis                           & Larval Ciona brain connectome            & Depth of synaptic contacts    \\
college\_freshmen                       & Friendship network Dutch freshmen        & Ratings -1 to +3              \\
copenhagen\_calls                       & Phone calls among students               & Call duration or count        \\
foodweb\_baywet                         & Everglades food web                      & Carbon flow                   \\
freshmen\_t0                            & Friendship network Groningen freshmen T0 & Friendship scale 1–5          \\
freshmen\_t2                            & Friendship network Groningen freshmen T2 & Friendship scale 1–5          \\
freshmen\_t3                            & Friendship network Groningen freshmen T3 & Friendship scale 1–5          \\
freshmen\_t5                            & Friendship network Groningen freshmen T5 & Friendship scale 1–5          \\
freshmen\_t6                            & Friendship network Groningen freshmen T6 & Friendship scale 1–5          \\
game\_thrones                           & Character co-appearance network          & Number of co-appearances      \\
highschool                              & High school friendship network           & Friendship named in surveys   \\
kangaroo                                & Dominance network of kangaroos           & Dominance frequency           \\
lesmis                                  & Les Misérables character network         & Scene co-appearances          \\
macaques                                & Dominance network of macaques            & Dominance frequency           \\
moreno\_sheep                           & Dominance network of sheep               & Dominance frequency           \\
plant\_pol\_kato                        & Plants linked to pollinators             & Visit frequency               \\
product\_space\_HS                      & Products linked by joint exports         & Proximity score               \\
product\_space\_SITC                    & Products linked by co-exports            & Co-export similarity          \\
psi                                     & Photosystem I chromophore network        & FRET efficiency               \\
residence\_hall                         & Friendship network in residence hall     & Friendship score 1–5          \\
rhesus\_monkey                          & Rhesus monkey grooming network           & Grooming frequency            \\
sp\_baboons\_observational              & Guinea baboon contact network            & Duration or frequency         \\
train\_terrorists                       & Madrid train bombing network             & Connection type (4 levels)    \\
unicodelang                             & Languages linked to countries            & Literate population share     \\
webkb\_cornell\_cocite           & Co-citation network Cornell WebKB        & Shared citations              \\
webkb\_cornell\_link1            & Cornell WebKB hyperlink network          & Number of hyperlinks          \\
webkb\_texas\_cocite             & Co-citation network Texas WebKB          & Shared citations              \\
webkb\_texas\_link1              & Texas WebKB hyperlink network            & Number of hyperlinks          \\
webkb\_washington\_cocite        & Co-citation network Washington WebKB     & Shared citations              \\
webkb\_washington\_link1         & Washington WebKB hyperlink network       & Number of hyperlinks          \\
webkb\_wisconsin\_cocite         & Co-citation network Wisconsin WebKB      & Shared citations              \\
webkb\_wisconsin\_link1          & Wisconsin WebKB hyperlink network        & Number of hyperlinks          \\
windsurfers                             & Social network of windsurfers            & Perceived closeness           
\end{tabular}
\end{ruledtabular}
\end{table*}

\subsection{Weighted Chung-Lu model}
\label{subsec:wchunglu}
In practice, sampling from the WCM requires creating and randomly matching stubs corresponding to all unit-weight edges. As the scale factor $A$ increases, the number of stubs grows proportionally, and the computational cost becomes prohibitive. To overcome this limitation, we adopt the natural extension of the Chung-Lu model~\cite{chung02} to the weighted case. Instead of explicitly generating stub matchings, we describe the distribution of edge weights directly: for off-diagonal pairs $(i\neq j)$ this is captured by a hypergeometric distribution, as we show below. A separate correction is required for diagonal entries $(i=j)$, since a self-loop is produced by pairings of stubs from the same vertex.

\paragraph{Off-diagonal entries $(i\neq j)$ via hypergeometric sampling.}
Let the strength sequence be $\{s_1,\ldots,s_n\}$ and $2W=\sum_i s_i$ the total number of stubs. We fix a vertex $i$ and consider the set $S_i$ of stubs that will be matched to $i$, via $s_i$ stubs; by construction $|S_i|=s_i$. In the random matching process, all size-$s_i$ subsets of the other $2W-1$ stubs are equally likely. The random weight $W_{ij}$, corresponding to the number of ($i$, $j$) connections, is therefore the number of times a stub of $i$ attaches to one of the $s_j$ stubs of $j$.
Since we are sampling $n=s_i$ items without replacement from a population of size $N=2W-1$ that contains $K=s_j$ successes (stubs of $j$), the distribution of $w_{ij}$ is hypergeometric:

\begin{equation}
w_{ij} \sim \mathrm{Hypergeometric}\big(N=2W-1,\;K=s_j,\;n=s_i\big).
\label{eq:hypergeometric}
\end{equation}
This yields the standard moments
\begin{align}
\mathbb{E}[w_{ij}] &= \frac{s_i s_j}{2W-1},
\\[3pt]
\noalign{\vskip3pt\hrule\vskip3pt}
\mathrm{Var}(w_{ij}) &=
   \frac{s_i\,\frac{s_j}{2W-1}\left(1-\frac{s_j}{2W-1}\right)\,(2W-1-s_i)}{2W-2}.
\label{eq:hypergeometricmoments}
\end{align}
Hence the expected strengths are preserved.

When all weights in the original network are scaled by a factor $A$, the strengths also scale as $s_i' = A s_i$ and the total number of stubs becomes $2W' = 2A W$. By replacing these values into Eq.~\ref{eq:hypergeometricmoments} we see that the variance grows linearly with $A$, like the mean:

\begin{equation}
\mathbb{E}[w_{ij}'] \;\approx\; A\,\frac{s_i s_j}{2W},
\qquad
\mathrm{Var}(w_{ij}') \;\approx\; A\,\frac{s_i s_j}{2W},
\end{equation}
up to negligible $O(1/A)$ corrections. Thus, when $A$ is large and the network remains sparse in terms of units of weight ($s_i, s_j \ll W$), the hypergeometric distribution converges to a Poisson with mean $\lambda_{ij} = A s_i s_j /(2W)$. In this regime, the classic Chung-Lu formulation is recovered as an asymptotic limit.

\paragraph{Self-loop model ($i=j$).}
When $i=j$, the hypergeometric argument above does not apply: a self-loop forms when two stubs of $i$ are paired to each other. Let $\mathcal{P}_i$ be the set of unordered pairs among $i$'s stubs, $|\mathcal{P}_i|=\binom{s_i}{2}$. For any specific pair $\{a,b\}\in\mathcal{P}_i$, the probability that $a$ matches to $b$ (and thus forms a loop) is $1/(2W-1)$, because $a$ chooses uniformly among the other $2W-1$ stubs. Therefore the expected value of the weight of the self-loop of $i$ is

\begin{equation}
\mathbb{E}[w_{ii}] \;=\; \binom{s_i}{2}\,\frac{1}{2W-1}
\;=\; \frac{s_i(s_i-1)}{2(2W-1)}\approx\frac{s_i^2}{4W}.
\end{equation}

The distribution of $w_{ii}$ is more complex~\cite{radicchi10}, because events for different pairs are not independent (a stub cannot participate in two loops). We approximate $w_{ii}$ by a binomial count of disjoint within-$i$ pairings:

\begin{equation}
w_{ii} \;\sim\; \mathrm{Binomial}\!\Big(\big\lfloor \tfrac{s_i}{2}\big\rfloor,\; p_i\Big),
\qquad
p_i \;=\; \frac{s_i-1}{2W-1}.
\label{eq:binomialselfloops}
\end{equation}

This choice is motivated by four considerations: (i) self-loops are pairs of stubs, not single draws; (ii) the binomial mean $\,\lfloor s_i/2\rfloor\,p_i$ matches $\mathbb{E}[w_{ii}]$ for even $s_i$ and is asymptotically exact when $W$ is large; (iii) by construction $w_{ii}\le \lfloor s_i/2\rfloor$, avoiding the overcounting that a naive hypergeometric for $i=j$ would permit; and (iv) for large $W$ the negative correlations among candidate pairs are $O(1/W)$, so treating disjoint pairings as approximately independent is accurate and yields the right variance scale. While this approach has a high computational cost, it only needs to be used a single time for each vertex.

In Appendix~\ref{appendix:weighted_chunglu_model} we illustrate that this is a good approximation of the actual distribution of weights on self-loops according to the WCM.


%

\subsection{CReMb variant and modularity compatibility}
\label{subsec:SECM}
Our variant of the CReMb relies on the fact that the baseline term $s_i s_j/2W$ in the modularity formulation (Eq.~\ref{eq:modul}) can be factorized into a structural component depending on degrees and a weighted component depending on strengths. In particular, modularity can be rewritten as
\begin{equation}
Q=\frac{1}{2W}\sum_{ij} \left[W_{ij}-\left(\frac{k_ik_j}{2m}\right)\left(\frac{s_is_j}{2W}\frac{2m}{k_i k_j}\right)\right]\delta_{g_ig_j},
\end{equation}
where $m$ is the number of edges in the network. This factorization naturally separates topology from weights in the construction of null models. Accordingly, in our model the adjacency matrix $B^{R}_{ij}$ is generated using the traditional configuration model, so that degrees match the empirical sequence. However, there is a chance of forming multi-edges, so the actual number of neighbors of each vertex may not be preserved.
Independently, weights are assigned through
\begin{equation}
S^{R}_{ij} \sim \text{Exponential}(\lambda),
\qquad
\lambda = \frac{W k_i k_j}{m s_i s_j},
\end{equation}
so that expected strengths coincide with the empirical sequence. For the cases in which there is a multi-edge between two vertices, we assign a weight value to each elementary edge: the weight of the multi-edge is the sum of the weights of the elementary edges.

The weight matrix of the null model is then obtained by the element-wise product
\begin{equation}
\mat{W}^{R} = \mat{B}^{R} \circ \mat{S}^{R}.
\end{equation}

This dual construction preserves the factorized structure of the modularity expectation, while ensuring that both degree and strength constraints are satisfied in expectation. As such, it represents a valid null model for weighted modularity and provides a practical way to compute network randomizations by independently sampling adjacency and weight contributions.

\subsection{Calculation of key variables}
To evaluate the statistical significance of network properties, we focused on three widely used variables of weighted networks: maximum modularity, maximum eigenvector centrality, and the weighted clustering coefficient of Onnela {\it et al.}~\cite{onnela05}. For each of them, we adopted standard implementations in widely used software packages, as detailed below.

Maximizing modularity is known to be an NP-hard problem~\cite{brandes2008modularity}, so computing the exact maximum is infeasible for most real-world networks. A more practical approach is to approximate the optimum using heuristic algorithms. Here, the maximum modularity was obtained by optimizing the weighted modularity function (Eq.~\ref{eq:modul}) with the Leiden algorithm~\cite{traag19}, as implemented in the \texttt{leidenalg} Python package. The Leiden algorithm is based on three iterative processes: (1) local movement of vertices to improve modularity, (2) refinement of the resulting communities to ensure they are internally well connected, and (3) aggregation of the network into a reduced graph on which the process is repeated. This iterative refinement yields high-modularity partitions that are structurally well defined while remaining computationally efficient for large networks.

The weighted eigenvector centrality was computed using the \texttt{igraph} Python package. The implementation relies on the ARPACK library, which estimates the leading eigenvector of the weight matrix through the implicitly restarted Arnoldi iteration method~\cite{lehoucq1998arpack}. This approach is efficient for large sparse matrices and yields the principal eigenvector \(\vec{c}\) from which vertex centralities are derived. In our analysis we considered the maximum eigenvector centrality, i.e., the largest entry of the leading eigenvector. We then L2-normalize this eigenvector to unit length,
\begin{equation}
\hat{\vec{c}} = \frac{\vec{c}}{\|\vec{c}\|_2},
\quad \text{where} \quad
\|\vec{c}\|_2 = \sqrt{\sum_i c_i^2}.
\end{equation}
All reported values (including the maximum entry) refer to the normalized vector \(\hat{\vec{c}}\).

Finally, the weighted clustering coefficient of Onnela et al.~\cite{onnela05} was calculated using the implementation available in the \texttt{networkx} package.

\subsection*{Acknowledgement} \label{sec:ack}
We would like to thank Olaf Sporns and Maria Grazia Puxeddu for pointing us to the data and the analysis that let us discover the scale dependence of the WCM. Also, we are grateful to Ginestra Bianconi, Diego Garlaschelli, Mariangeles Serrano, and Tiziano Squartini for valuable feedback on the manuscript.
This project was partially supported by the Air Force Office of Scientific Research under award numbers FA9550-19-1-0391, FA9550-21-1-0446 and FA9550-24-1-0039, and by the National Science Foundation under award numbers 1927418, and by the National Institutes of Health under awards U01 AG072177 and U19 AG074879. This work utilized Indiana University Jetstream2 CPU through allocation BIO230158 and CIS230183 from the Advanced Cyber-infrastructure Coordination Ecosystem: Services \& Support (ACCESS) program, which is supported by National Science Foundation grants \#2138259, \#2138286, \#2138307, \#2137603, and \#2138296.

\subsection*{Contributions}
All authors conceived the research, discussed, and wrote the manuscript. F. N. S. and S.K. performed the analysis and experiments.

\subsection*{Competing interest}
The authors have no competing interest.

\subsection*{Data sharing plans}
We made available the code and documentations to reproduce all results. See \url{https://github.com/filipinascimento/wmodularity} for details.

\clearpage

\appendix

\section{Weighted Chung-Lu Model Validation}
\label{appendix:weighted_chunglu_model}
\begin{figure*}
    \centering
    \includegraphics[width=0.90\textwidth]{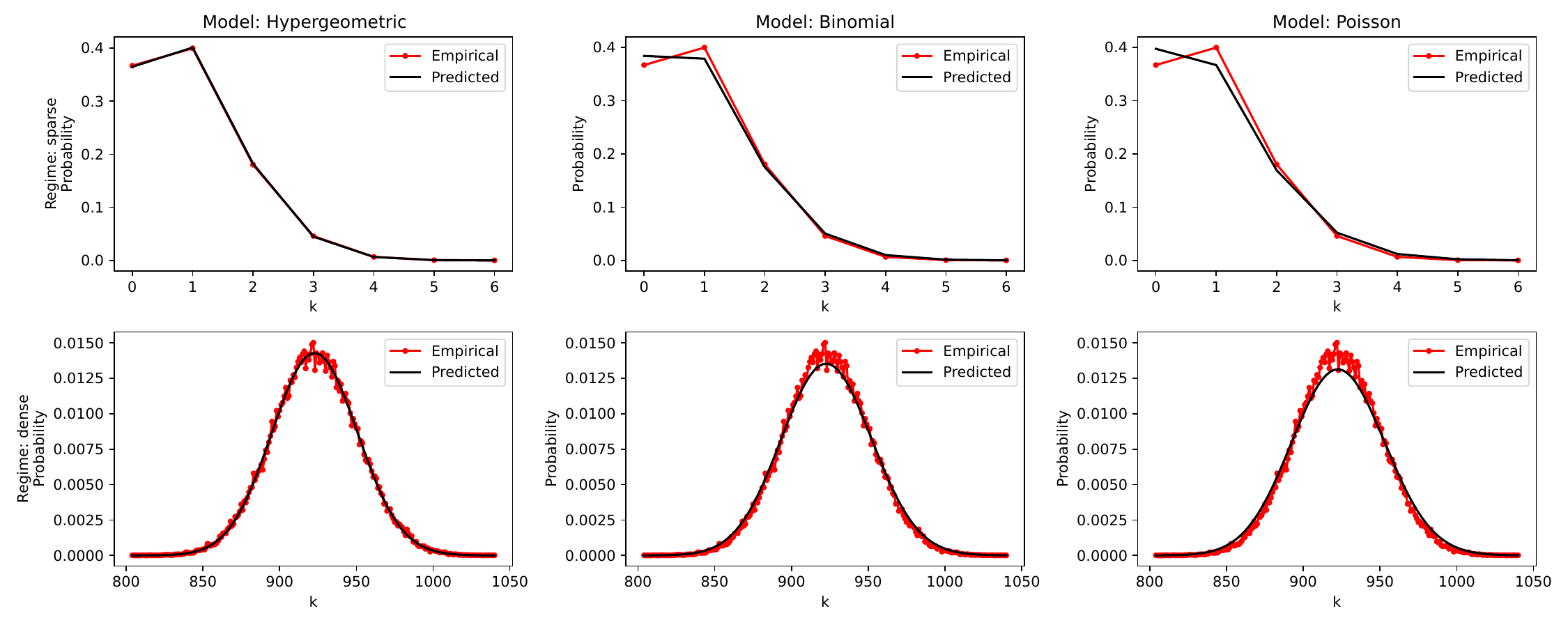}
    \caption{Empirical versus predicted distributions of edge weights in the configuration model for the Zachary Karate Club network. The distributions correspond to the total weight of a single edge (between nodes 9 and 16) measured across multiple realizations. Red markers represent the empirical probabilities obtained from simulations, while black lines show the analytical predictions based on the weighted Chung–Lu approximation using different weight-generating distributions (hypergeometric, binomial, and Poisson). The top row shows the sparse regime ($A = 1$), while the bottom row shows the dense regime ($A = 1000$), where $A$ is a scaling factor applied to the original node strengths.}
    \label{fig:model_correspondence_different}
\end{figure*}

\begin{figure*}
    \centering
    \includegraphics[width=0.90\textwidth]{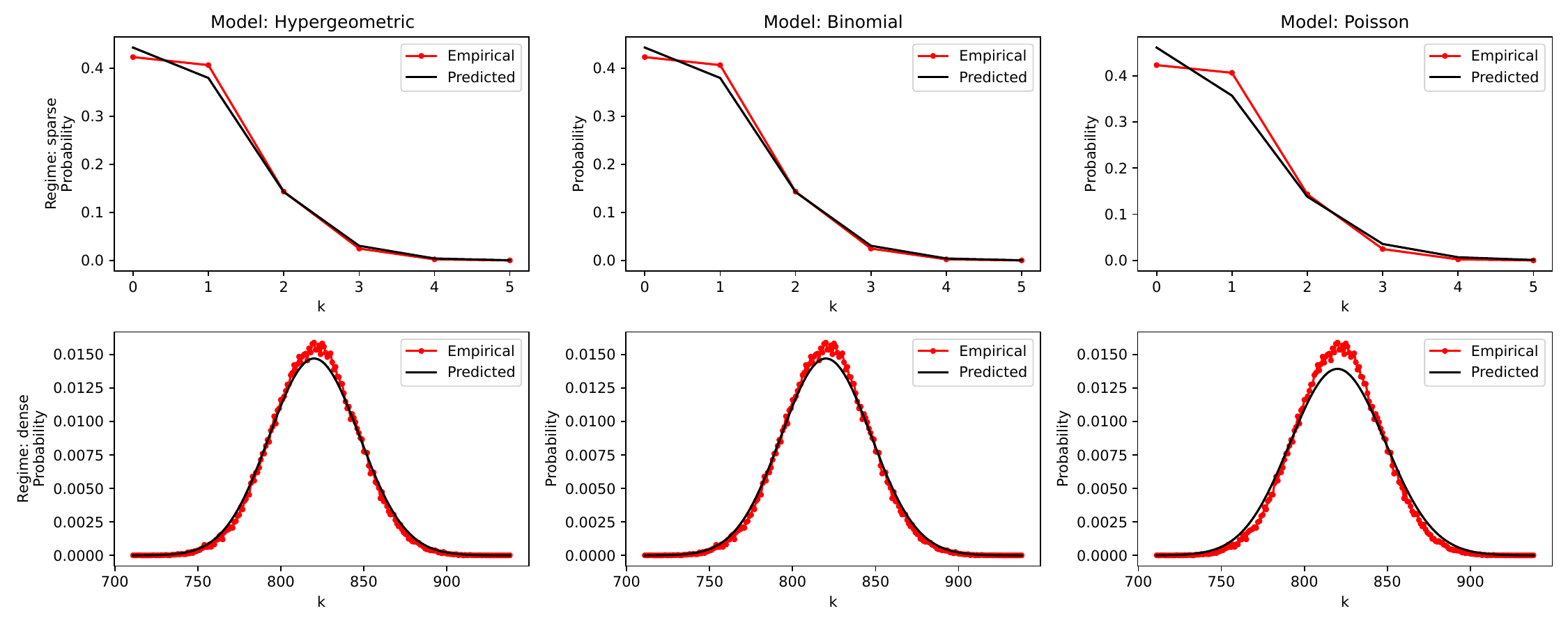}
    \caption{Same as Fig.~\ref{fig:model_correspondence_different}, but for the self-loop corresponding to node 16.}
    \label{fig:model_correspondence_self}
\end{figure*}

Figures~\ref{fig:model_correspondence_different} and~\ref{fig:model_correspondence_self} illustrate the comparison between the empirical edge weight distributions obtained from multiple realizations of the configuration model and the corresponding analytical predictions from the weighted Chung–Lu approximation. The different panels show the behavior across regimes of increasing density, controlled by the scaling factor $A$. As summarized in Tables~\ref{tab:model_comp_A1}–\ref{tab:model_comp_A10000}, the hypergeometric variant provides the best overall agreement with the empirical results, yielding the smallest $L_1$ distances and KL divergences in the sparse regime ($A=1$). However, as the network becomes denser, the differences between the hypergeometric, binomial, and Poisson formulations gradually vanish, and in the very dense limit ($A=10000$), all models converge to nearly identical outcomes.

\begin{table}[]
\centering
\caption{Comparison between empirical and predicted weight distributions across all edges in the Zachary Karate Club network. Each row reports the mean and standard deviation of the $L_1$ distance and Kullback–Leibler (KL) divergence between the empirical weight distributions obtained from WCM simulations and the analytical predictions based on the Weighted Chung–Lu approximation with different weight-generating distributions (hypergeometric, binomial, and Poisson). Results correspond to the sparse regime ($A=1$), as in Fig.~\ref{fig:model_correspondence_different}.}
\label{tab:model_comp_A1}
\begin{ruledtabular}
\begin{tabular}{lcc}
Model & $L_1$ mean $\pm$ std & KL mean $\pm$ std \\
  \colrule
Hypergeometric & 0.012 $\pm$ 0.006 & 0.0003 $\pm$ 0.0002 \\
Binomial         & 0.054 $\pm$ 0.009 & 0.0026 $\pm$ 0.0007 \\
Poisson          & 0.089 $\pm$ 0.009 & 0.006 $\pm$ 0.001
\end{tabular}
\end{ruledtabular}
\end{table}

\begin{table}[]
\centering
\caption{Same as Table~\ref{tab:model_comp_A1}, but for the dense regime ($A=1000$).}
\label{tab:model_comp_A1000}
\begin{ruledtabular}
\begin{tabular}{lcc}
Model & $L_1$ mean $\pm$ std & KL mean $\pm$ std \\
  \colrule
Hypergeometric & 0.093 $\pm$ 0.007 & 0.011 $\pm$ 0.001 \\
Binomial         & 0.104 $\pm$ 0.006 & 0.014 $\pm$ 0.001 \\
Poisson          & 0.119 $\pm$ 0.007 & 0.017 $\pm$ 0.001 \\
\end{tabular}
\end{ruledtabular}
\end{table}

\begin{table}[]
\centering
\caption{Same as Table~\ref{tab:model_comp_A1}, but for the very dense regime ($A=10000$).}
\label{tab:model_comp_A10000}

\begin{ruledtabular}
\begin{tabular}{lcc}
Model & $L_1$ mean $\pm$ std & KL mean $\pm$ std \\
  \colrule
Hypergeometric & 0.51 $\pm$ 0.02 & 0.26 $\pm$ 0.02 \\
Binomial         & 0.52 $\pm$ 0.02 & 0.27 $\pm$ 0.02 \\
Poisson          & 0.52 $\pm$ 0.02 & 0.27 $\pm$ 0.02 \\
\end{tabular}
\end{ruledtabular}
\end{table}

\bibliographystyle{unsrt}
\bibliography{unit_invariance}

\end{document}